\documentclass [11pt,a4paper] {article}

\usepackage[cp1252]{inputenc}

\usepackage{amssymb}
\usepackage{amsmath}
\usepackage{amsfonts,amssymb}

\usepackage[dvips]{graphicx}

\usepackage{bbm}

\DeclareMathAlphabet{\mathpzc}{OT1}{pzc}{m}{it}

\begin{document}

\title{Intuitionistic interpretation of quantum mechanics}

\author{Arkady Bolotin\footnote{$Email: arkadyv@bgu.ac.il$\vspace{5pt}} \\ \textit{Ben-Gurion University of the Negev, Beersheba (Israel)}}

\maketitle

\begin{abstract}\noindent In the present paper, the decision problem of the Schr\"odinger equation (asking whether or not a given Hamiltonian operator has the nonempty solution set) is represented as a logical statement. As it is shown in the paper, the law of excluded middle would be applicable to the introduced statement if and only if quantum fundamentalism (asserting that everything in the universe is ultimately describable in quantum-mechanical terms) held. But, since the decision problem of the Schr\"odinger equation is in general undecidable, such a statement is allowed to be other than true or false, explicitly, it may fail to have truth values at all. This makes possible to abandon the law of excluded middle together with quantum fundamentalism in the proposed intuitionistic interpretation of quantum mechanics.\\

\end{abstract}

\section{Introduction}\label{Introduction}

\noindent There is a striking similarity between intuitionistic logic (i.e., a form of constructivism – a philosophy of mathematics characterized by the requirement that when a mathematical object is asserted to exist, positive evidence witnessing its truth should be given) and the Copenhagen interpretation of quantum mechanics.\\

\noindent Really, unlike classical logic, where either a statement $p$ or its negation $\neg p$ must be true regardless of whether one has positive evidence for either $p$ or $\neg p$, in intuitionistic logic, statements are assigned a truth value only if they can be given a direct proof. Therewithal, unproved statements do not have any intermediate logical value so that their relation to truth remains unknown until they are either proved or disproved.$\,$\footnote{\label{f1}Introductory texts on intuitionistic logic and constructivism can be found, e.g., in \cite{Bezhanishvili, Bridges06, Artemov}.\vspace{5pt}}\\

\noindent Now, compare that to the Copenhagen-based explanation of \textit{the quantum interference experiment}, in which an electron beam passing through a plate pierced by two parallel symmetrically positioned slits is observed on a screen behind the plate. According to this explanation, a sentence like ``before being absorbed at the screen an electron went through either slit 1 or slit 2'' has no meaning unless an experiment is performed that decides through which particular slit the electron goes.$\,$\footnote{\label{f2}Here it is assumed that there is no other way the electron emitted from the source can hit the screen except going through the slits.\vspace{5pt}} Only when there is a piece of apparatus that is capable of deciding which statement is true: \textit{the statement} $p$ that the electron goes through slit 1 or \textit{the statement} $\neg p$ that the electron goes through slit 2, it is permissible to say the electron went through either slit. Without such an apparatus, i.e., without positive evidence, one may not assign a truth value to the statements $p$ and $\neg p$. If one, nevertheless, does this, i.e., if one assumes that since \textit{the “middle” statement is excluded by logic}, the disjunction $p\vee \neg p$ is always true, and starts to make any deductions from that assumption, errors will be made in the analysis. In Feynman's words, “This is the logical [to be more specific, \textit{intuitionistic } -- A.B.] tightrope on which we must walk if we wish to describe nature successfully.”\cite{Feynman}\\

\noindent Likewise, the Copenhagen interpretation leaves open the nature of the electron between the source and the screen, so that it cannot ask itself the question about whether this nature is point-like or not-point-like. This means that between the source and the screen statements like ``the electron is a particle'' (statement $p$) and ``the electron is a wave'' (statement $\neg p$) are unproved and hence remain of unknown truth value. And only by placing detectors directly behind the slits (in this way getting positive evidence supporting the correctness of either $p$ or $\neg p$ and as a result the truthfulness of $p\vee \neg p$) these statements can be valuated.\\

\noindent The attentive reader may enquire is there more to this similarity between intuitionism and Copenhagen interpretation than just a coincidence? Is it, for example, possible that quantum phenomena appear mysterious to us only because we try to apply to them the law of the excluded middle? If so, puzzles of quantum mechanics might be dissolved once intuitionistic logic was adopted in quantum theory.\\

\noindent The problem, however, is that the mathematical formalism of quantum mechanics sooner demonstrates the failure of the distributive law of propositional logic than the breakdown of the law of the excluded middle, and in this way quantum formalism would rather advocate the adoption of quantum logic than intuitionistic logic.$\,$\footnote{\label{f3}But comparing intuitionistic logic with the traditional quantum logic (of Birkhoff and von Neumann), one can easily notice the superiority of the former. As it has been elegantly put in the paper \cite{Caspers}, quantum logic is ``too radical in giving up distributivity (for one thing rendering it problematic is to interpret the logical operations $\wedge$ and $\vee$ as conjunction and disjunction, respectively)'' at the same time as it is ``not radical enough in keeping the law of excluded middle (so that it falls victim to Schr\"odinger's cat and the like)''.\vspace{5pt}} As a result, it is still unclear whether a consistent intuitionistic (or any constructivist for that matter) interpretation of quantum mechanics is possible.\\

\noindent Although attempts have been made to analyze different aspects of the foundations of quantum mechanics within Bishop's constructive mathematics (equivalent to mathematics done with intuitionistic logic) and to find constructive substitutes of quantum formalism (see, for example, \cite{Bridges99, Richman, Bridges00}), those substitutes have never become sufficient for developing a full intuitionistic counterpart of quantum mechanics.\\

\noindent Strictly speaking, the task of developing such a counterpart involves finding a correspondence between logical propositions and mathematical objects of quantum formalism such that these mathematical objects together form the structure of a Heyting algebra, i.e., the formal algebraic structure of intuitionistic logic.$\,$\footnote{\label{f4}A Heyting algebra is a bounded lattice, i.e., an algebraic structure of the form $(L,\wedge,\vee,\top,\bot)$ consisting of a set $L$, two binary operations $\wedge$ and $\vee$, and the top and bottom elements denoted as $\top$ (``true'') and $\bot$ (``false''), respectively. A Heyting algebra is a proper generalization of the notion of a Boolean algebra, namely, every Boolean algebra is a Heyting algebra and a Heyting algebra is Boolean if and only if $\neg\neg\ p=p$ for all $p \in L$ \cite{Borceux}.\vspace{5pt}}\\

\noindent Usually in quantum physics, a proposition is written in the form $(A\in \triangle a)$, where $a\in \mathbb{R}$ is a value of an observable $A$ and $\triangle a$ is a subset of $\mathbb{R}$, so that $(A\in \triangle a)$ can be understood as the assertion that the value of $A$ is in the subset $\triangle a$. Accordingly, in the event of an actual measurement of the observable $A$, the proposition $(A\in \triangle a)$ may become determined affirmatively and, correspondingly, its negation $\neg(A\in \triangle a)\equiv(A\not\in \triangle a)$ may be determined negatively. However, in the case of a measurement of another, incompatible observable $B$ (such that $AB\not=BA$ ), the logical disjunction $(A\in \triangle a)\vee \neg(A\in \triangle a)$ will have no meaning, that is, it will be neither true nor false (until the subsequent measurement of the observable $A$ will prove it).\\

\noindent Thus, what it is considered true strongly depends on the measurement context. However, propositions are specified without referring to the measurement context (given that such an external information is not contained in the standard quantum formalism). So, this could be seen as an explanation of why it is not possible to unambiguously associate propositions with subsets of the Hilbert space \cite{Hermens}.\\

\noindent As it is shown in \cite{Caspers}, if one adopts a richer structure, which is capable of accounting for measuring contexts, like a topos (an algebraic structure in category theory that behaves much like the category \textbf{Sets} of sets except that all proofs in it have to be constructive \cite{Heunen}), then it becomes possible to construct a Heyting algebra to obtain an intuitionistic quantum logic for systems associated with a finite-dimensional Hilbert space. But even granting that, not only is the topos-theoretical approach a very abstract one from a mathematical point of view, but it also brings about propositions that have no meaning from a physical point of view. In consequence, it is difficult to find a physical interpretation of the logic obtained by such an approach.\\

\noindent So, in the present paper, another, completely different way for `constructivization' of quantum mechanics is proposed. Namely, to formulate an intuitionistic interpretation of quantum mechanics it is proposed to revise the assumption of the universal validity of Schr\"odinger dynamics adopted in quantum mechanics (specifically, in the decoherent-based versions of the Copenhagen interpretation).\\

\section{The decision problem of the Schr\"odinger equation}\label{Two}

\noindent Let $\left|\!\left.{\mathit{u}} \!\right.\right\rangle$ -- an element of an abstract (complex) vector space $\mathcal{U}$ -- represent the exact solution$\,$\footnote{\label{f5}Even though a general, mathematically rigorous definition of ``exact solution'' is difficult enough to outline (since such a definition should, on the one hand, possess a certain heuristic value giving one a chance to apply a mathematical formalism but, on the other hand, have a clear physical meaning – see on this subject, for example, \cite{Turbiner}), for the purpose of this paper, by ``the exact solution $\left|\!\left.{\mathit{u}} \!\right.\right\rangle$'' one can understand either the analytic (algebraic) expression for $\left|\!\left.{\mathit{u}} \!\right.\right\rangle$ explicitly presented in terms of some previously known functions or the exact algorithm for calculating numerical values of $\left|\!\left.{\mathit{u}} \!\right.\right\rangle$ to an arbitrary precision.\vspace{5pt}} to the general, time-dependent Schr\"odinger equation describing how the quantum state of a physical system $S$ changes with time\\

\begin{equation} \label{1} 
        i\hbar\frac{\partial}{\partial t} \!
        \left|\!\left. {\mathit{u}} \!\right.\right\rangle
        =
        H_S \! \left|\!\left. {\mathit{u}} \!\right.\right\rangle
   \;\;\;\;  ,
\end{equation}
\smallskip

\noindent where $H_S$ symbolizes the Hamiltonian operator identified with the system $S$, while time $t$ is a parameter ranging over the set of all real numbers $\mathbb{R}$. The general equation (\ref{1}) can be viewed as \textit{the function problem} formulated as follows:\\

\begin{quote}
\noindent \textit{Given an explicit expression for the Hamiltonian operator $H_S$, find all solutions $\left|\!\left. {\mathit{u}} \!\right.\right\rangle$ to the equation resulted from the insertion of this expression into the general Schr\"odinger equation or decide that no such solution exists.}\\
\end{quote}

\noindent The corresponding \textit{decision problem of the Schr\"odinger equation} can be then defined as the logical statement ${\Pi}_{\mathit{u}}(H_S)$, which is taken to be true if the solution set $\{\left|\!\left. {\mathit{u}} \!\right.\right\rangle\}$, i.e., the set of all $\left|\!\left. {\mathit{u}} \!\right.\right\rangle$ for which the Schr\"odinger equation with the particular Hamiltonian operator $H_S$ holds, is not empty and false otherwise, i.e.$\,$\footnote{\label{f6}In this paper, ``statements'' and ``propositions'' are synonymous and so they are used interchangeably, despite the fact that from the point of view of philosophy and theoretical linguistics there is some distinction between them. Particularly, propositions admit multiple representations, while statements are their representations \cite{Quine}.\vspace{5pt}}\\

\begin{equation} \label{2} 
        {\Pi}_{\mathit{u}}(H_S)
        \equiv
        \left(
        \left|
        \left\{
        \left|\!\left. {\mathit{u}} \!\right.\right\rangle \in \mathcal{U}
        \,\Big|\,
        i\hbar\frac{\partial}{\partial t} \!
        \left|\!\left. {\mathit{u}} \!\right.\right\rangle
        =
        H_S \! \left|\!\left. {\mathit{u}} \!\right.\right\rangle
        \right\}
        \right| >0
        \right)
   \;\;\;\;  .
\end{equation}
\smallskip

\noindent Let the domain of discourse of all allowable Hamiltonian operators $H_S$ be denoted by $\mathbb{H}$, then the assertion of the universality of Schr\"odinger dynamics can be abbreviated as the following statement:\\

\begin{equation} \label{3} 
        \forall H_S \in \mathbb{H} \,
        \big(
        {\Pi}_{\mathit{u}}(H_S)
        \big)
   \;\;\;\;  .
\end{equation}
\smallskip

\noindent For example, this statement will be affirmatively determined if there is a witness, which provides a proof that ${\Pi}_{\mathit{u}}(H_S)$ is true given any allowable Hamiltonian operator $H_S$.$\,$\footnote{\label{f7}Here and hereinafter, the Brouwer-Heyting-Kolmogorov (BHK) interpretation of logical connectives and quantifiers is used \cite{Taranovsky}.\vspace{5pt}}\\

\noindent Let $\left|\!\left. {\psi} \!\right.\right\rangle$, $\left|\!\left. {\phi} \!\right.\right\rangle$ and $\left|\!\left. {\varphi} \!\right.\right\rangle$ be system's pure arbitrary quantum states represented by vectors (state vectors) in the Hilbert space ${\mathcal{H}}_S$ associated with the system $S$. Consider the following claim:\\

\begin{quote}
\noindent \textit{The system in any pure quantum state $\left|\!\left. {\psi} \!\right.\right\rangle$ can be regarded as being partly in each of two other pure quantum states $\left|\!\left. {\phi} \!\right.\right\rangle$ and $\left|\!\left. {\varphi} \!\right.\right\rangle$ such that $\left|\!\left. {\psi} \!\right.\right\rangle=\alpha\left|\!\left. {\phi} \!\right.\right\rangle+\beta\left|\!\left. {\varphi} \!\right.\right\rangle$ where the complex coefficients $\alpha$ and $\beta$ must satisfy the normalization condition ${\alpha}^2+{\beta}^2=1$.}\\
\end{quote}

\noindent This claim can be shorten to the equivalent denoted by $C_{\psi}({\mathcal{H}}_S)$:\\

\begin{equation} \label{4} 
        C_{\psi}({\mathcal{H}}_S)
        \equiv
        \forall \left|\!\left. {\psi} \!\right.\right\rangle\! ,  
                 \left|\!\left. {\phi} \!\right.\right\rangle\! , 
                 \left|\!\left. {\varphi} \!\right.\right\rangle
        \in {\mathcal{H}}_S \:\:
        \forall \alpha,  
                 \beta 
        \in \mathbb{C} \,
        \Big(
           \left|\!\left. {\psi} \!\right.\right\rangle
           =
           \alpha\left|\!\left. {\phi} \!\right.\right\rangle+\beta\left|\!\left. {\varphi} \!\right.\right\rangle
           \,\wedge\,
           {\alpha}^2+{\beta}^2=1
        \Big)
   \;\;\;\;  .
\end{equation}
\smallskip

\noindent Suppose the domain of discourse specifying all possible Hilbert spaces ${\mathcal{H}}_S$, which are associated with physical systems, to be referred to as ${\mathcal{H}}$ (this domain ${\mathcal{H}}$ can be thought of as a subset of abstract complete inner product vector spaces that meet physical constraints); then, the statement of the universality of the quantum superposition principle can be expressed as follows:\\

\begin{equation} \label{5} 
        \forall {\mathcal{H}}_S \in {\mathcal{H}} \,
        \big(
        C_{\psi}({\mathcal{H}}_S)
        \big)
   \;\;\;\;  .
\end{equation}
\smallskip

\noindent This statement will have a definite true value, say ``true'', if there is a witness which proves the truthfulness of $C_{\psi}({\mathcal{H}}_S)$ for any allowed Hilbert space ${\mathcal{H}}_S$.\\

\noindent Using the expressions (\ref{3}) and (\ref{5}), \textit{the assumption of the universal applicability of quantum mechanics} can be formulated as the logical conjunction\\

\begin{equation} \label{6} 
        A_Q
        \equiv
        \forall H_S \in \mathbb{H} \:\:
        \forall {\mathcal{H}}_S \in {\mathcal{H}} \,
        \Big(
        {\Pi}_{\mathit{u}}(H_S)
           \,\wedge\,
        C_{\psi}({\mathcal{H}}_S)
        \Big)
   \;\;\;\;   
\end{equation}
\smallskip

\noindent that is true when there is a witness, which, given any arbitrary physical system $S$ specified in the form of its Hamiltonian operator $H_S$ and its associated Hilbert space ${\mathcal{H}}_S$ as input, provides positive evidence supporting both ${\Pi}_{\mathit{u}}(H_S)$ and $C_{\psi}({\mathcal{H}}_S)$ (i.e., validity of both Schr\"odinger dynamics and the quantum superposition principle for this system). Oppositely, to prove the negation $\neg A_Q$ of this assumption, i.e.\\

\begin{equation} \label{7} 
        \neg A_Q
        =
        \exists H_S \in \mathbb{H} \:\:
        \exists {\mathcal{H}}_S \in {\mathcal{H}} \,
        \Big(
        \neg {\Pi}_{\mathit{u}}(H_S)
           \,\vee\,
        \neg C_{\psi}({\mathcal{H}}_S)
        \Big)
   \;\;\;\;  ,
\end{equation}
\smallskip

\noindent a witness demonstrating contradictory of either ${\Pi}_{\mathit{u}}(H_S)$ or $C_{\psi}({\mathcal{H}}_S)$ must be provided for at least one physical system $S$.\\

\noindent In contrast with the constructivist point of view (according to which knowing that a statement is true means having a proof of it), in the modern versions of the Copenhagen interpretation, i.e., ones that include decoherent-based explanations of the appearance of wavefunction collapse, it is assumed – notwithstanding the absence of direct proof – that the formalism of quantum mechanics can be extrapolated from the level of single electrons and atoms (where one actually can find satisfactory evidence that this formalism is quantitatively valid) up to the level of macroscopic systems (such as observers and their measurement devices) and even up to the level of the universe as a whole.\\

\noindent Therewith, the motive for such an extrapolation is this: As there is no experimental evidence that quantum mechanics is not valid in the regions where it has not been directly tested, the principle of Occam's razor would certainly suggest that the most intellectually economical attitude in this situation is to consider that the assumption of the universal applicability of quantum mechanics $A_Q$ is true \cite{Leggett}.\\

\noindent But in fact, this `intellectually economical' attitude mirrors the proof by contradiction of classical logic. Indeed, according to the reasoning behind the extrapolation of quantum formalism, the lack of evidence that the assumption $A_Q$ is wrong must imply that the assumption $A_Q$ is true, which can be formulated as \textit{the double negation elimination}, i.e.\\

\begin{equation} \label{8} 
        \neg\neg A_Q
        \rightarrow
         A_Q
   \;\;\;\;  .
\end{equation}
\smallskip

\noindent Recall that unlike ${\Pi}_{\mathit{u}}(H_S)$, the statement $C_{\psi}({\mathcal{H}}_S)$ has no existential components in its definition$\,$\footnote{\label{f8}Bearing in mind that the existential quantifier symbol $\exists$ can be interpreted as ``it is a set'', the statement ${\Pi}_{\mathit{u}}(H_S)$ is logically equivalent to the assertion $\exists \left|\!\left. {\mathit{u}} \!\right.\right\rangle \in \mathcal{U}\, P(\left|\!\left. {\mathit{u}} \!\right.\right\rangle\!,t,H_S)$, where $P(\left|\!\left. {\mathit{u}} \!\right.\right\rangle\!,t,H_S)$ is the predicate ``$i\hbar\frac{\partial}{\partial t} \!  \left|\!\left. {\mathit{u}} \!\right.\right\rangle=H_S \! \left|\!\left. {\mathit{u}} \!\right.\right\rangle$''.\vspace{5pt}} and so the witness to $C_{\psi}({\mathcal{H}}_S)$ has nothing \textit{decidedly nontrivial} to do, which means that $C_{\psi}({\mathcal{H}}_S)$ can be treated classically: $\neg\neg C_{\psi}({\mathcal{H}}_S) \rightarrow C_{\psi}({\mathcal{H}}_S)$. Consequently, in order to prove the negation elimination (\ref{8}), the proof of the following negation elimination is necessary\\

\begin{equation} \label{9} 
        \forall H_S \in \mathbb{H} \:\:
        \Big(
        \neg\neg {\Pi}_{\mathit{u}}(H_S)
        \rightarrow
        {\Pi}_{\mathit{u}}(H_S)
        \Big)
   \;\;\;\;  .
\end{equation}
\smallskip

\noindent However, the latter holds only if the decision problem of the Schr\"odinger equation is decidable, that is to say, if the statement ${\Pi}_{\mathit{u}}(H_S)$ can be assigned a definite truth value (one of the two element set $\mathbb{B}=\{\top,\bot\}$) such that the following statement of excluded middle \textit{always holds true}:\\

\begin{equation} \label{10} 
        \forall H_S \in \mathbb{H} \:\:
        \Big(
        {\Pi}_{\mathit{u}}(H_S)
        \vee
        \neg {\Pi}_{\mathit{u}}(H_S)
        \Big)
        =
        \top
   \;\;\;\;  .
\end{equation}
\smallskip

\noindent In its turn, this statement can be proved only if the decision problem of the Schr\"odinger equation can be solved by an effective decision procedure, i.e., by an algorithm.\\

\noindent At the moment, besides \textit{a brute-force method} (aka \textit{exhaustive search} or \textit{direct search} that sequentially tests each possibility – i.e., a candidate solution – in order to determine whether or not it belongs to the solution set $\{\left|\!\left. {\mathit{u}} \!\right.\right\rangle\}$), there is no other procedure, which can solve the decision problem of the Schr\"odinger equation for \textit{any allowable Hamiltonian} operator $H_S$.$\,$\footnote{\label{f9}See, for example, the book \cite{Popelier} that focuses on non-mainstream methods to solve the molecular electronic Schr\"odinger equation and discusses whether a truly predictive computational scheme other than built on brute force would be ever possible in quantum chemistry.\vspace{5pt}}\\

\noindent As to brute-force, however, even though this general method can always find the correct solution set $\{\left|\!\left. {\mathit{u}} \!\right.\right\rangle\}$ or prove that it is empty, its computational cost (e.g., a number of steps required to do so) is proportional to the number of candidate solutions for $\{\left|\!\left. {\mathit{u}} \!\right.\right\rangle\}$ which tends to grow very quickly as the size of the list of amplitudes describing the quantum state of the system increases. So, when this list is infinite (just as in the case of a particle on a lattice that can be in any one of infinitely many discrete positions), the computational cost of the brute-force method may rise to an infinity.\\

\noindent As a result, there does not exist a single (i.e., a generic) effective method that can in a finite amount of steps correctly solve the decision problem of the Schr\"odinger equation for all allowable Hamiltonian operators $H_S$. Thus, one must admit (at least in the present state of the knowledge) that the statement ${\Pi}_{\mathit{u}}(H_S)$ is in general undecidable.$\,$\footnote{\label{f10}The decision problem of the Schr\"odinger equation is parallel to \textit{the general spectral gap problem} (the central one in quantum many-body physics) that asks whether or not a given Hamiltonian operator $H_S$ has a spectral gap, i.e. the energy difference between the ground state and the first excited state of the system in the thermodynamic limit. Truly, if there were a generic algorithm capable of obtaining the solution set $\{\left|\!\left. {\mathit{u}} \!\right.\right\rangle\}$ of the Schr\"odinger equation for any allowable Hamiltonian operator $H_S$, then such an algorithm would be able to not only valuate the statement ${\Pi}_{\mathit{u}}(H_S)$, but also answer the question whether the spectrum of the resultant eigenvalues is discrete and gapped or continuous and gapless. On the other hand, as it is demonstrated in the paper \cite{Cubitt}, when the allowable expressions for the Hamiltonian operator $H_S$ in the Schr\"odinger equation are restricted to translationally invariant ones given by nearest neighbor interactions on a 2D lattice, there cannot exist any procedure which, given the matrix elements of the local interactions of the Hamiltonian $H_S$, determines whether the resulting spectrum of states is gapped or gapless. This implies that the general spectral gap problem is undecidable. But what is more, one can infer from this that the generic algorithm for solving the decision problem of the Schr\"odinger equation cannot exist as well.\vspace{5pt}}\\

\noindent Even so, the decoherent-based interpretations of quantum mechanics are based upon the supposition declaring\\

\begin{quote}
\noindent \textit{Schr\"odinger dynamics is universally valid,}\\
\end{quote}

\noindent which can be written down in formal terms as the following propositional formula

\begin{equation} \label{11} 
        \forall H_S \in \mathbb{H} \:\:
        {\Pi}_{\mathit{u}}(H_S)
        =
        \top
   \;\;\;\;  ,
\end{equation}
\smallskip

\noindent where symbol ``='' defines the notion of a valuation; specifically, the formula (\ref{11}) assigns the truth value $\top$ (or the meaning of ``true'') to every statement ${\Pi}_{\mathit{u}}(H_S)$ in the entire domain $\mathbb{H}$ of allowable Hamiltonian operators $H_S$. This propositional formula is logically equivalent to the postulate asserting\\

\begin{quote}
\noindent \textit{All Schr\"odinger equations are exactly solvable.}\\
\end{quote}

\noindent On the other hand, in view of the linearity of the Schr\"odinger equation, the formula (11) implies that superpositions of quantum states of any physical system – including, for example, the system consisting of the silver atom plus the measuring apparatus in the Stern-Gerlach experiment – can be solutions of the Schr\"odinger equation. As a result, there ought to be superpositions in the measurement outcomes, i.e., at the macroscopic level (for example, there should be a superposition state of the atom plus apparatus in which apparatus pointers are in various positions at the same time). Clearly, this contradicts ordinary experience which is the essence of the (in)famous measurement problem of quantum mechanics.\\

\noindent As it is well known, decoherence per se does not constitute an adequate answer to the measurement problem,$\,$\footnote{\label{f11}Decoherence is only a way to show why no macroscopic superposed state can be observed, whereas the quantum entanglement between the atom, the apparatus and the environment never disappears. See, for example, \cite{Schlosshauer04, Zinkernagel11, Zwirn} for the discussion concerning decoherence and further references.\vspace{5pt}} therefore, responses to this problem have been sought by invoking additional assumptions such as observer-induced collapse, many worlds, hidden variables, modified dynamics, transactional ``handshakes'', modal interpretations, just to name a few. However, explaining why no quantum strangeness is seen in the measurement results all those assumptions clearly or arguably go beyond the standard quantum formalism. More importantly, none of them is widely believed to be free of accompanying conceptual problems \cite{Tammaro}.\\

\noindent So, in order to resolve the measurement problem it seems necessary to refuse to recognize the statement of excluded middle ${\Pi}_{\mathit{u}}(H_S)\!\vee\!\neg{\Pi}_{\mathit{u}}(H_S)$ as being always true, which causes this problem in the first place.\\

\noindent Really, by treating the statement of the universal validity of Schr\"odinger dynamics as a truthful one – which is only possible if ${\Pi}_{\mathit{u}}(H_S)\!\vee\!\neg{\Pi}_{\mathit{u}}(H_S)$ is a logical truth – one endows macroscopic systems with new features such as macroscopic quantum superposition states. And upon doing so, one creates a paradox, which requires new theories for its explanation.\\

\section{The intuitionistic interpretation of quantum mechanics}\label{Three}

\noindent But instead, let us consider an intuitionistic interpretation of quantum mechanics, in which the statement of excluded middle ${\Pi}_{\mathit{u}}(H_S)\!\vee\!\neg{\Pi}_{\mathit{u}}(H_S)$ can still be proved for some systems, however, this statement does not hold universally (as it does in the decoherent-based Copenhagen interpretations).\\

\noindent More specifically, let us consider an interpretation of quantum mechanics, which is based on intuitionistic (constructive) logic and the supposition that there is no generic solution to the decision problem of the Schr\"odinger equation working in all cases of the Hamiltonian operator $H_S$ (despite the fact that there may be solutions that work in some cases). Accordingly, for an arbitrary physical system $S$, the statement ${\Pi}_{\mathit{u}}(H_S)$ will be neither true nor false unless one has a proof of this statement.\\

\subsection{The main assumption}\label{Three_one}

\noindent The main assumption of the proposed intuitionistic interpretation can be expressed in the form of the following assertion:\\

\begin{quote}
\noindent \textit{Schr\"odinger dynamics is possibly valid.}\\
\end{quote}

\noindent With the modal operator of possibility $\Diamond$, this assertion can be formally written down as the formula\\

\begin{equation} \label{12} 
        \Diamond {\Pi}_{\mathit{u}}(H_S)
        =
        \top
   \;\;\;\;  ,
\end{equation}
\smallskip

\noindent which assigns the truth value $\top$ to the proposition ``it is possible that ${\Pi}_{\mathit{u}}(H_S)$''.$\,$\footnote{\label{f12}Many aspects of modal logic are covered in \cite{Borghini}.\vspace{5pt}} This assumption is logically equivalent to the postulate asserting\\

\begin{quote}
\noindent \textit{There might be exactly solvable Schr\"odinger equations.}\\
\end{quote}

\noindent As a consequence of the assumption (\ref{12}), it follows that there may be physical systems for which the formula ${\Pi}_{\mathit{u}}(H_S)\!\vee\!\neg{\Pi}_{\mathit{u}}(H_S)$ is not true. Indeed, using the modal operator of necessity $\Box$ and the analogy of de Morgan's laws $\neg\Diamond p\leftrightarrow \Box\neg p$, one can write the negation of the assumption (\ref{12}) as the following propositional expression\\

\begin{equation} \label{13} 
        \Box\neg {\Pi}_{\mathit{u}}(H_S)
        =
        \bot
   \;\;\;\;  ,
\end{equation}
\smallskip

\noindent which states\\

\begin{quote}
\noindent \textit{Not-Schr\"odinger dynamics is impossible (i.e., not possibly valid).}\\
\end{quote}

\noindent Making use of the characteristic axiom of modal logic $\left(\Box\neg p=\bot\right) \rightarrow \left(\neg p=\bot\right)$ (that reads ``if not-$p$ is impossible, then not-$p$ is not the case'' \cite{Chellas}), one can get that the logical disjunction on values of two statements $\Diamond {\Pi}_{\mathit{u}}(H_S)$ and $\neg{\Pi}_{\mathit{u}}(H_S)$ is always true:\\

\begin{equation} \label{14} 
        \Diamond {\Pi}_{\mathit{u}}(H_S)
        \vee
        \neg {\Pi}_{\mathit{u}}(H_S)
        =
        \top
   \;\;\;\;  .
\end{equation}
\smallskip

\noindent Now, let us go back to the assumption (\ref{12}): Assigning the ``true” value to \textit{the possibility} that the statement ${\Pi}_{\mathit{u}}(H_S)$ is true makes ${\Pi}_{\mathit{u}}(H_S)$ contingent because its actual truth value remains unsettled.$\,$\footnote{\label{f13}Since the negation of the statement ${\Pi}_{\mathit{u}}(H_S)$ has no existential components (see footnote \ref{f8}), it can be treated classically: in particular, the negation $\neg{\Pi}_{\mathit{u}}(H_S)$ is assumed to be false. But despite that, the statement ${\Pi}_{\mathit{u}}(H_S)$ itself cannot be asserted since it has not been proved yet.\vspace{5pt}} Surely, since ${\Pi}_{\mathit{u}}(H_S) \rightarrow \Diamond {\Pi}_{\mathit{u}}(H_S)$ and not vice versa, the valuation $\Diamond {\Pi}_{\mathit{u}}(H_S)=\top$ does not entail the truthfulness of ${\Pi}_{\mathit{u}}(H_S)$.$\,$\footnote{\label{f14}Such a difference is strictly due to the accepted supposition that there does not exist a generic algorithm for solving the decision problem of the Schr\"odinger equation. If this problem were universally solvable, one would have ${\Pi}_{\mathit{u}}(H_S) \rightarrow \Diamond {\Pi}_{\mathit{u}}(H_S)$ and $\Diamond {\Pi}_{\mathit{u}}(H_S) \rightarrow {\Pi}_{\mathit{u}}(H_S)$.\vspace{5pt}} In this way, the disjunction (\ref{14}) indicates that for some systems the law of excluded middle would not hold, i.e.:\\

\begin{equation} \label{15} 
        \exists H_S \in \mathbb{H} \:\:
        \Big(
        {\Pi}_{\mathit{u}}(H_S)
        \vee
        \neg {\Pi}_{\mathit{u}}(H_S)
        \Big)
        \not=
        \top
   \;\;\;\;   
\end{equation}
\smallskip

\noindent (where the existential introduction $p(H_{S^{\prime}}) \rightarrow \exists H_S\! \in\! \mathbb{H} \,\, p(H_{S})$ is assumed). From a mathematical point of view, this attests an intuitionistic logic of the proposed interpretation of quantum mechanics.$\,$\footnote{\label{f15}Formally speaking, this means that the statement ${\Pi}_{\mathit{u}}(H_S)$ is an element of a Heyting algebra.\vspace{5pt}}\\

\noindent Thus, contrary to the decoherent-based interpretations of quantum mechanics, the proposed interpretation limits the applicability of the quantum mechanical description \textit{only to those systems for which the statement ${\Pi}_{\mathit{u}}(H_S)$ can be affirmed explicitly}, that is, can be valuated positively by an arithmetical or recursive procedure in a finite number of steps, i.e., by an algorithm.$\,$\footnote{\label{f16}Yet, in some cases, the statement ${\Pi}_{\mathit{u}}(H_S)$ can be proved without an algorithm that solves it. According to the Church-Turing thesis (CTT), any function that is computable is computable by some Turing machine \cite{Gandy}; hence, the statement ${\Pi}_{\mathit{u}}(H_S)$ is provable when it is Turing-computable. On the other hand, according to the Physical CTT, any physical process (i.e., anything doable by a physical system) is computable by some Turing machine \cite{Piccinini}. Thus, a witness, which provides a proof that ${\Pi}_{\mathit{u}}(H_S)$ is true, can be physically implemented i.e., realized physically. In other words, this witness can be a physical process (e.g., the Stern-Gerlach experiment) empirically demonstrating that Schr\"odinger dynamics is valid for a particular system.\vspace{5pt}}\\

\subsection{Turing-computability}\label{Three_two}

\noindent From a computational theoretical perspective, the statement ${\Pi}_{\mathit{u}}(H_S)$ \textit{might be Turing-computable} if the Hilbert space ${\mathcal{H}}_S$ associated with the system $S$ were finite-dimensional or it were capable of being safely approximated by a finite-dimensional vector space ${\mathbb{C}}^N \simeq \{\left|\!\left. {n} \!\right.\right\rangle\}^{N}_{n=1} \subset {\mathcal{H}}_S$.\\

\noindent In more detail, an arbitrary pure quantum state $\left|\!\left. {\Psi} \!\right.\right\rangle$ of the system $S$ can be presented by the equation\\

\begin{equation} \label{16} 
        \left|\!\left. {\Psi} \!\right.\right\rangle
        =
        \sum^{N}_{n=1}  {\Psi}_n \! 
                                               \left|\!\left. n \!\right.\right\rangle
        +
        \mathbbm{X}_{\infty} \! \left|\!\left. {\Psi} \!\right.\right\rangle          
   \;\;\;\;  ,
\end{equation}
\smallskip

\noindent where $\{{\Psi}_n \equiv \langle n\! \left|\!\left. {\Psi} \!\right.\right\rangle\}^N_{n=1}$ is the list of complex amplitudes describing the quantum state $\left|\!\left. {\Psi} \!\right.\right\rangle$ and $\mathbbm{X}_{\infty}$ is the projector onto the reminder of the system's infinite Hilbert space ${\mathcal{H}}_S$, that is, on the relative complement ${\mathcal{H}}_S \setminus {\mathbb{C}}^N$. In the case, in which the component $\mathbbm{X}_{\infty} \! \left|\!\left. {\Psi} \!\right.\right\rangle$ could be safely neglected, the number of candidate solutions for the solution set $\{\left|\!\left. {\mathit{u}} \!\right.\right\rangle\}$ might be limited. This suggests that in this case the statement ${\Pi}_{\mathit{u}}(H_S)$ might be proved in a finite number of steps by at least a brute-force method.$\,$\footnote{\label{f17}This could be infeasible to achieve in practice, though, due to an exponential amount of steps required by the brute force method.\vspace{5pt}}\\

\noindent Then again, if the replacement of an infinite Hilbert space ${\mathcal{H}}_S$ by a finite dimensional vector space ${\mathbb{C}}^N$ were impossible to justify, then the statement ${\Pi}_{\mathit{u}}(H_S)$ would be undecidable which would make the quantum mechanical description of the system $S$ inapplicable.\\

\noindent Take, for example a typical microscopic system consisting of one or a few isolated subatomic particles moving at less than a relativistic velocity in a confined area and involving only limited energies (similar to energies of electrons in an atom or a solid). For this system, one can safely assume a discrete spectrum of energies, which is upper limited by some finite level, and in that way one can restrict the orthonormal basis of the infinite Hilbert space to some truncated basis $\{\left|\!\left. {n} \!\right.\right\rangle\}^{N}_{n=1}$. Because of this, for such a system, the decision problem of the Schr\"odinger equation has the possibility of being solvable (e.g., by a brute force method) and consequently the quantum mechanical description may be applicable.$\,$\footnote{\label{f18}What’s more, for this (i.e., typical microscopic) system, quantum formalism turns out to be constructive (i.e., computable): Namely, in the truncated basis $\{\left|\!\left. {n} \!\right.\right\rangle\}^{N}_{n=1}$ one would need only a finite number of steps in order to construct (to any given precision) the matrix ${\mathsf{H}}_S=\left(H_{nn}\right)^N_{nn}$ describing the Hamiltonian operator $H_S$ in a numerical (computer) representation or to exhibit the wavefunction $\Psi(x)=\langle x\! \left|\!\left. {\Psi} \!\right.\right\rangle$ of the system.\vspace{5pt}}\\

\noindent In opposition, a typical macroscopic system always interacts (even only weakly) with a large part of the universe, or perhaps, with all of it. Therefore, most of the system’s degrees of freedom – in particular the macroscopic ones – can vary continuously (i.e., in an unbroken series or pattern) and in an unconfined, unbounded manner. Following this line of reasoning, one can remark that any discretization of the system’s macroscopic degrees of freedom (in order to perform numerical computations) would correspond to an infinite-dimensional Hilbert space for which a replacement by a finite-dimensional numerical basis may not be vindicated. As a consequence, the decision problem of the Schr\"odinger equation for a typical macroscopic system could not be solved by a brute force method (as it would fail to halt). This implies that in the given case the statement ${\Pi}_{\mathit{u}}(H_S)$ would be undecidable, i.e., it would remain neither true nor false, making in this way the quantum mechanical description of a typical macroscopic system irrelevant.$\,$\footnote{\label{f19}Furthermore, due to the equivalence of \textit{boundedness} and \textit{continuity}, unbounded linear Hermitian operators in an infinite Hilbert space are discontinuous and thus generate Turing-noncomputability. This infers that the application of quantum formalism to a typical macroscopic system would necessarily cause the existence of Turing-noncomputable (and so nonconstructive) mathematical entities in the description of the system which cannot be regarded as experimentally testable. See \cite{Hellman} and \cite{Myrvold94, Myrvold95} for the discussion concerning constructivism and computability in quantum mechanics.\vspace{5pt}}\\

\subsection{Heisenberg's cut}\label{Three_three}

\noindent So, a Heisenberg's cut – i.e., an imagined demarcation line between the quantum mechanical and classical descriptions – is determined in the proposed interpretation by the Turing computability of the decision problem of the Schr\"odinger equation. This may explain the basic difficulty that quantum mechanics meets in locating the ``shifty split'' between quantum and classical phenomena.\\

\noindent First, neither mass, nor geometrical size nor any other physical characteristic can serve by itself as a qualitative criterion for the Turing computability. Next, there is a liberty in choosing the truncated basis $\{\left|\!\left. {n} \!\right.\right\rangle\}^{N}_{n=1}$ (that approximates a computable representation of a system in which the statement ${\Pi}_{\mathit{u}}(H_S)$ might be provable/decidable) for an infinite-dimensional Hilbert space. Furthermore, any change to this truncated basis – for example, by the inclusion of a few more variables into the quantum mechanical description of the system – can bring about a shift in the Heisenberg’s cut.\\

\subsection{The von Neumann measurement scheme}\label{Three_four}

\noindent The fact that the quantum mechanical description of an arbitrary macroscopic system could not be considered constructive invalidates the von Neumann quantum measurement scheme.\\

\noindent As stated by the assumptions on which the von Neumann scheme is based, the measurement process is governed by the Schr\"odinger equation for the Hamiltonian operator $H_{S+M}$ describing the composite quantum system $S+M$ which contains the measured microscopic system $S$ and the measuring macroscopic apparatus $M$ devised to measure the observable $A$ of the microscopic system $S$ (in order to facilitate this, the apparatus $M$ supposedly has the ready-state $\left|\!\left. {M_0} \!\right.\right\rangle$ and the set of mutually orthogonal states $\{\left|\!\left. {M_n} \!\right.\right\rangle\}$, all orthogonal to $\left|\!\left. {M_0} \!\right.\right\rangle$, corresponding to different macroscopic configurations of $M$ similar to different positions of a pointer along a scale). The linearity of the Schr\"odinger equation yields the eigenvector-eigenvalue link – the perfect correlation between the initial (i.e., at the moment $t=0$ preceding the measurement) state $\left|\!\left. {a_n} \!\right.\right\rangle$ of the microsystem $S$ and the final (i.e., at the moment $t$ in the last part of the measurement) state $\left|\!\left. {M_n} \!\right.\right\rangle$ of the apparatus $M$, i.e.\\

\begin{equation} \label{17} 
        \left|\!\left. {a_n} \!\right.\right\rangle \! \left|\!\left. {M_0} \!\right.\right\rangle 
        \: \longrightarrow \:
        \left|\!\left. {\Psi}_t \!\right.\right\rangle
        =
        \left|\!\left. {a_n} \!\right.\right\rangle  \! \left|\!\left. {M_n} \!\right.\right\rangle 
    \;\;\;\;   
\end{equation}
\smallskip

\noindent in such a way that if the final state of the apparatus is $\left|\!\left. {M_n} \!\right.\right\rangle$, then the observable $A$ has the value $a_n$.\\

\noindent When the initial state of the microscopic system $S$ is just one of the vectors $\left|\!\left. {a_n} \!\right.\right\rangle$ like in Eq.\ref{17}, the proposition\\

\begin{equation} \label{18} 
        \left(A=a_n\right) 
        \equiv
        \left(
        \left|\!\left. {a_n} \!\right.\right\rangle  \! \left|\!\left. {M_n} \!\right.\right\rangle
        =
        e^\frac{-itH_{S+M}}{\hbar}
        \left|\!\left. {a_n} \!\right.\right\rangle  \! \left|\!\left. {M_0} \!\right.\right\rangle
        \right) 
    \;\;\;\;   
\end{equation}
\smallskip

\noindent contains no logical connectives such as $\vee$, $\rightarrow$ and $\exists$, and therefore it demands no decidedly nontrivial witness to its truthfulness. This means that the proposition $\left(A=a_n\right)$ can be treated classically, particularly, $\left(A=a_n\right)\vee\left(A\not=a_n\right)$ can be considered as an instance of the law of excluded middle.\\

\noindent However, in the situation, in which the initial state of the microsystem $S$ is a superposition of the vectors like $\left|\!\left. {a_n} \!\right.\right\rangle$, for example, $\left|\!\left. {n+k} \!\right.\right\rangle=\frac{1}{\sqrt{2}}\left(\left|\!\left. {a_n} \!\right.\right\rangle + \left|\!\left. {a_k} \!\right.\right\rangle\right)$, and, as a consequence, the final state of the system $S+M$ is given by the macroscopic superposition\\

\begin{equation} \label{19} 
        \frac{1}{\sqrt{2}}
        \left(\left|\!\left. {a_n} \!\right.\right\rangle + \left|\!\left. {a_k} \!\right.\right\rangle\right) 
        \! \left|\!\left. {M_0} \!\right.\right\rangle 
        \: \longrightarrow \:
        \left|\!\left. {\Psi}_t \!\right.\right\rangle
        =
        \frac{1}{\sqrt{2}}
        \left(
        \left|\!\left. {a_n} \!\right.\right\rangle \! \left|\!\left. {M_n} \!\right.\right\rangle
        +
        \left|\!\left. {a_k} \!\right.\right\rangle \! \left|\!\left. {M_k} \!\right.\right\rangle
        \right) 
    \;\;\;\;  ,
\end{equation}
\smallskip

\noindent the law of excluded middle is not applicable, which constitutes a contradiction of the von Neumann scheme that is built on the law of excluded middle.\\

\noindent Indeed, let us assume (for the sake of simplicity) that the observable $A$ can only take on two values $a_n$, $a_k$ and consider the proposition $\left(A\in \{a_n,a_k\}\right)$ asserting that there is a final state $\left|\!\left. {\Psi}_t \!\right.\right\rangle$ of the microsystem $S$ + the apparatus $M$, in which the observable $A$ has a definite value, i.e., either $a_n$ or $a_k$:\\

\begin{equation} \label{20} 
        \left(A\in \{a_n,a_k\}\right)
        \equiv
        \left(
           \left|\!\left. {\Psi}_t \!\right.\right\rangle
           =
          \left|\!\left. {a_n} \!\right.\right\rangle \! \left|\!\left. {M_n} \!\right.\right\rangle
          \vee
          \left|\!\left. {\Psi}_t \!\right.\right\rangle
           =
          \left|\!\left. {a_k} \!\right.\right\rangle \! \left|\!\left. {M_k} \!\right.\right\rangle
        \,\Big|\,
          \left|\!\left. {\Psi}_t \!\right.\right\rangle
           =
           e^\frac{-itH_{S+M}}{\hbar}
          \left|\!\left. {n+k} \!\right.\right\rangle  \! \left|\!\left. {M_0} \!\right.\right\rangle
        \right)
   \;\;\;\;  .
\end{equation}
\smallskip

\noindent According to classical logic (on which any measuring macroscopic apparatus is based), the proposition $\left(A\in \{a_n,a_k\}\right)=\left(A=a_n\right)\vee\left(A=a_k\right)$ must be a logical truth.$\,$\footnote{\label{f20}Here it is suggested that $\neg\left(A=a_n\right)=\left(A=a_k\right)$ and correspondingly $\neg\left(A=a_k\right)=\left(A=a_n\right)$.\vspace{5pt}} But this contradicts to the scheme of dynamical evolution presented in Eq.\ref{19}, according to which the proposition $\left(A\in \{a_n,a_k\}\right)$ must be false: This scheme declares that the observable $A$ may simultaneously  have both values $a_n$ and $a_k$, that is, $\left(A=a_n\right)\wedge\left(A=a_k\right)=\top$.\\

\noindent The standard way out from this contradiction (while maintaining classical logic) is to supplement the Schr\"odinger equation with \textit{the wave-packet reduction postulate}, which states that at the end of the measurement process the final state vector $\left|\!\left. {\Psi}_t \!\right.\right\rangle$ reduces to one of its terms: either $\left|\!\left. {a_n} \!\right.\right\rangle \! \left|\!\left. {M_n} \!\right.\right\rangle$ or $\left|\!\left. {a_k} \!\right.\right\rangle \! \left|\!\left. {M_k} \!\right.\right\rangle$ (thus making the proposition $\left(A\in \{a_n,a_k\}\right)$ true).$\,$\footnote{\label{f21}In the decoherent-based interpretations, the interaction with the environment produces essentially a randomization of the phases associated to the different components of the final state vector, a process which can be seen as an apparent collapse of this vector into one of these components \cite{Zeh}.\vspace{5pt}}\\

\noindent By contrast, intuitionistic logic (adopted in the proposed interpretation of quantum mechanics) requires that if the proposition $\left(A\in \{a_n,a_k\}\right)$ is asserted to be true, then a witness proving its truthfulness must be given since it contains the logic operator $\vee$ of disjunction.$\,$\footnote{\label{f22}Besides, this proposition can be rewritten so that it will contain the existential quantification $\exists$ together with the predicate variable $\left|\!\left. {\Psi}_t \!\right.\right\rangle$, which is the solution to the Schr\"odinger equation for the Hamiltonian $H_{S+M}$: $\left(A\in \{a_n,a_k\}\right)\equiv\exists\left|\!\left. {\Psi}_t \!\right.\right\rangle \in \{\left|\!\left. {a_n} \!\right.\right\rangle \! \left|\!\left. {M_n} \!\right.\right\rangle,\left|\!\left. {a_k} \!\right.\right\rangle \! \left|\!\left. {M_k} \!\right.\right\rangle\}\, P(\left|\!\left. {\Psi}_t \!\right.\right\rangle)$, where $P(\left|\!\left. {\Psi}_t \!\right.\right\rangle)$ is the predicate enclosing this equation.\vspace{5pt}} As it can be seen from Eq.\ref{20}, such a witness can be a solution to the Schr\"odinger equation for the Hamiltonian $H_{S+M}$ when the input $\left|\!\left. {\Psi}_0 \!\right.\right\rangle =\left|\!\left. {n+k} \!\right.\right\rangle  \! \left|\!\left. {M_0} \!\right.\right\rangle$ is given.\\

\noindent On the other hand, according to the proposed intuitionistic interpretation, Schr\"odinger dynamics is only possibly valid, and so $\Diamond {\Pi}_{\mathit{u}}(H_{M})=\top$, which implies that one has no guarantee that the Schr\"odinger equation for the Hamiltonian operator $H_M$ describing an arbitrary macroscopic apparatus $M$ could be exactly solved. So (unless one has empirical evidence proving the validity of Schr\"odinger dynamics for this particular apparatus $M$), the statement ${\Pi}_{\mathit{u}}(H_M)$ must stay on contingent.\\

\noindent The undecidability of the statement ${\Pi}_{\mathit{u}}(H_M)$ implies that the ready-state $\left|\!\left. {M_0} \!\right.\right\rangle \in {\mathbb{C}}^N \simeq \{\left|\!\left. {M_n}\!\right.\right\rangle\}^N_n$ where ${\mathbb{C}}^N$ is the vector space over the mutually orthogonal quantum states $\left|\!\left. {M_n} \!\right.\right\rangle$ of the apparatus $M$, could not be proven to exist. Hence, the predicate $P(\left|\!\left. {\Psi}_t \!\right.\right\rangle)$ of the proposition $\left(A\in \{a_n,a_k\}\right)$, which can be presented as\\

\begin{equation} \label{21} 
        P(\left|\!\left. {\Psi}_t \!\right.\right\rangle)
        \equiv
        \Big(
          \left|\!\left. {\Psi}_t \!\right.\right\rangle
           =
           e^\frac{-itH_{S+M}}{\hbar}
          \left|\!\left. {\Psi}_0 \!\right.\right\rangle
        \Big)
        \wedge
        \Big(\!
          \left|\!\left. {n+k} \!\right.\right\rangle
        \!\Big)
        \wedge
        \Big(\!
          \left|\!\left. {M_0} \!\right.\right\rangle
        \!\Big)
   \;\;\;\;  ,
\end{equation}
\smallskip

\noindent may not be algorithmically analyzed for its overall truth value since\\

\begin{equation} \label{22} 
        \Big(\!
          \left|\!\left. {M_0} \!\right.\right\rangle \in {\mathbb{C}}^N
        \!\Big)
        =
        \Big(\!
          {\Pi}_{\mathit{u}}(H_M)
        \!\Big)
        =
        \{\}
   \;\;\;\;  ,
\end{equation}
\smallskip

\noindent where the empty set $\{\}$ represents ``neither true nor false''. This means that the proposition $\left(A\in \{a_n,a_k\}\right)$ cannot be decided computationally, and so without direct evidence (i.e., without a result of the measurement of the observable $A$), one may not assign any truth value to the proposition $\left(A\in \{a_n,a_k\}\right)$.\\

\noindent So, the undecidability of the proposition $\left(A\in \{a_n,a_k\}\right)$ discards the necessity of the reduction postulate in the measurement scheme: In the proposed intuitionistic interpretation, this proposition \textit{does not transform from false to true during the measurement} in consequence of the wave-packet reduction, \textit{but merely remains neither false nor true} (in this way rendering invalid Eq.\ref{19}) till it is shown to be true by an actual measurement result at the end of the measurement process.$\,$\footnote{\label{f23}The asymmetry of the proposition $\left(A\in \{a_n,a_k\}\right)$ can be attributable to the asymmetry of the statement ${\Pi}_{\mathit{u}}(H_{(\cdot)})$. Really, let us consider the negation of the predicate $P(\left|\!\left. {\Psi}_t \!\right.\right\rangle)$: $\neg P(\left|\!\left. {\Psi}_t \!\right.\right\rangle)=(\neg ( \left|\!\left. {\Psi}_t \!\right.\right\rangle = e^{{(-itH_{S+M})}/{\hbar}} \left|\!\left. {\Psi}_0 \!\right.\right\rangle ) ) \vee \left(\neg \left(\left|\!\left. {n+k} \!\right.\right\rangle \right) \right)\vee \left(\neg {\Pi}_{\mathit{u}}(H_M) \right)$. According to the main assumption of the proposed interpretation, not-Schr\"odinger dynamics is impossible, which means that $\neg ( \left|\!\left. {\Psi}_t \!\right.\right\rangle \!=\! e^{{(-itH_{S+M})}/{\hbar}} \left|\!\left. {\Psi}_0 \!\right.\right\rangle )\!=\!\bot$ and also $\neg {\Pi}_{\mathit{u}}(H_M)\!=\!\bot$. But since $\neg \left(\left|\!\left. {n+k} \!\right.\right\rangle \right)\!=\!\bot$ (due to the initial condition of the considered problem), one finds $\neg P(\left|\!\left. {\Psi}_t \!\right.\right\rangle)\!=\!\bot$ and thus $\neg\left(A\in \{a_n,a_k\}\right)\!=\!\bot$. As a result, if $\left(A\in \{a_n,a_k\}\right)$ is decidable, that is, if $\left(A\in \{a_n,a_k\}\right)$ can be assigned a definite truth value, then $\left(A\in \{a_n,a_k\}\right)=\top$. One can observe from here that an actual measurement of the observable $A$ may only prove the truthfulness of the proposition $\left(A\in \{a_n,a_k\}\right)$.\vspace{5pt}}\\

\subsection{Counterfactuals in the intuitionistic interpretation}\label{Three_five}

\noindent The undecidability of the decision problem of the Schr\"odinger equation makes impossible to speak meaningfully of the definiteness of the results of measurements that have not been performed.\\

\noindent To show this, let us discuss the case in which two microsystems $S_A$ and $S_B$, whose states are determined by the observables $A$ and $B$, respectively, taking only on the values $a_1$, $a_2$ and $b_1$, $b_2$, are entangled before the measurement process so that at the moment $t=0$ the states of these microsystems are described by the entangled pure state $\left|\!\left. {\psi}_0 \!\right.\right\rangle$ (similar to one of four Bell states).\\

\noindent Let the composite (spatially separated bipartite) system governed by the Hamiltonian operator $H_C$ have a source that emits pairs of the microsystems $S_A$ and $S_B$ occupying the entangled state $\left|\!\left. {\psi}_0 \!\right.\right\rangle$ into two opposite directions, with the microsystem $S_A$ sent to destination where there is a macroscopic apparatus $M_A$ devised to measure the observable $A$, and the microsystem $S_B$ sent to the place of a macroscopic apparatus $M_B$ devised to measure the observable $B$.\\

\noindent Let us introduce the 2-tuple $\left(A,B\right)$ of pairs of values of the observables $A$ and $B$ and consider the following proposition\\

\begin{equation} \label{23} 
        \big(
           \left(A\in \{a,\neg a\}\right)
           \wedge
           \left(B\in \{b,\neg b\}\right)
        \big)
        =
        \big(
           (A,B) \in \{(a,b),(a,\neg b),(\neg a,b),(\neg a,\neg b)\}
        \big)
   \;\;\;\;  ,
\end{equation}
\smallskip

\noindent where the following abbreviations are used: $a=a_1$, $\neg a=a_2$ and $b=b_1$, $\neg b=b_2$. This proposition asserts that there is a final state of the considered composite system, in which the tuple $\left(A,B\right)$ has a definite value, that is:\\

\begin{equation} \label{24} 
        \big(
           \left(A\in \{a,\neg a\}\right)
           \wedge
           \left(B\in \{b,\neg b\}\right)
        \big)
        \!\equiv\!
        \left(
           \left|\!\left. {\Psi}_t \!\right.\right\rangle
           \!\in\!
          \{\left|\!\left. {\Psi}_i \!\right.\right\rangle\}^4_{i=1}
        \,\Big|\,
          \left|\!\left. {\Psi}_t \!\right.\right\rangle
           =
           e^\frac{-itH_{C}}{\hbar}
          \left|\!\left. {\psi}_0 \!\right.\right\rangle
            \! \left|\!\left. {M_{A0}} \!\right.\right\rangle
            \! \left|\!\left. {M_{B0}} \!\right.\right\rangle
        \right)
   \;\;\;\;  ,
\end{equation}
\smallskip

\noindent in which $\left|\!\left. {\Psi}_1 \!\right.\right\rangle=\left|\!\left. {a} \!\right.\right\rangle  \! \left|\!\left. {b} \!\right.\right\rangle  \! \left|\!\left. {M_{a}} \!\right.\right\rangle \! \left|\!\left. {M_{b}} \!\right.\right\rangle$, $\left|\!\left. {\Psi}_2 \!\right.\right\rangle=\left|\!\left. {a} \!\right.\right\rangle  \! \left|\!\left. {\neg b} \!\right.\right\rangle  \! \left|\!\left. {M_{a}} \!\right.\right\rangle \! \left|\!\left. {M_{\neg b}} \!\right.\right\rangle$, and so on.\\

\noindent Thus, if the observable $A$ were chosen to be measured and the proposition $(A=a)$ were shown to be true, then in any interpretation of quantum mechanics that adopts classical (or quantum) logic, Eq.\ref{23} would be as follows:

\begin{equation} \label{25} 
        \Big(
           \left(A=a\right)
           \wedge
           \left(B\in \{b,\neg b\}\right)
        \Big)
        =
        \Big(
           (A,B) \in \big\{(a,\{b,\neg b\}\big\}
        \Big)
        =
        \top
   \;\;\;\;  .
\end{equation}
\smallskip

\noindent This means that -- no matter how far apart from each other the macroscopic apparatuses $M_A$ and $M_B$ might be -- a counterfactual measurement of the observable $B$ must always be included together with the factual measurement of the observable $A$. In other words, in such interpretations the statistical population describing observable $A$ would contain two pairs $(a,b)$ and $(a,\neg b)$ for each possible value of $B$ regardless of whether or not an actual measurement of the observable $B$ has been performed.$\,$\footnote{\label{f24}Suppose, for example, that the entangled state $\left|\!\left. {\psi}_0 \!\right.\right\rangle$ of the microsystems $S_A$ and $S_B$ has been prepared so that the tuple $(A,B)$ can only contain the ordered pairs of values of the observables $A$ and $B$: $(A,B)\in\{(a,b),(\neg a,\neg b)\}$. Next, if the observable $A$ had been chosen to be measured and shown to be equal to $a$, i.e., $(A=a)=\top$, then in accordance with the interpretations based on classical or quantum logic, the result obtained by the spatially separated apparatus $M_B$ would be absolutely certain: $B=b$. And if at the place of the apparatus $M_B$ there was another macroscopic apparatus $M_C$ devised to measure the incompatible observable $C$ such that $CB\not=BC$, the outcome of the measurement would be not certain even without disturbing the microsystem $S_B$ at all.\vspace{5pt}}\\

\noindent In contrast to this, in the intuitionistic interpretation of quantum mechanics, the truth value of the proposition $\left((A\in \{a,\neg a\})\wedge (B\in \{b,\neg b\})\right)$ is computationally undecidable (as the statements ${\Pi}_{\mathit{u}}(H_{M_A})$ and ${\Pi}_{\mathit{u}}(H_{M_B})$ are both undecidable if the apparatuses $M_A$ and $M_B$ are both typical macroscopic systems). Consequently, this proposition could be assigned the truth value $\top$ only if \textit{there were witnesses for both $\left(A\in \{a,\neg a\}\right)$ and $\left(B\in \{b,\neg b\}\right)$}.\\

\noindent For example, assume, that the observable $A$ has been chosen to be measured and shown to be equal to $a$. This provides the witness for $(A=a)=\top$ but at the same time the proposition $\left(B\in \{b,\neg b\}\right)$ remains of unknown truth value, i.e., $\left(B\in \{b,\neg b\}\right)=\{\}$, and hence in the place of Eq.\ref{23} one would find

\begin{equation} \label{26} 
        \Big(
           \left(A=a\right)
           \wedge
           \{\}
        \Big)
        =
        \Big(
           (A,B) \in \big\{(a,\{\}\big\}
        \Big)
        =
        \{\}
   \;\;\;\;  ,
\end{equation}
\smallskip

\noindent which means that in the given case the statistical population describing observable $A$ may only contain the value $a$.$\,$\footnote{\label{f25}Therewith, a correlation between results of the distant measurements of the observables $A$ and $B$ can only be established when and where the data from both apparatuses $M_A$ and $M_B$ can be available.\vspace{5pt}}\\

\noindent Consequently, the proposed intuitionistic interpretation rejects the possibility of reliable measurements of the counterfactual and definite kind by making such a possibility \textit{logically invalid} for typical macroscopic systems.\\

\section{Discussion}\label{Four}

\noindent Let us very briefly examine possible arguments against the proposed intuitionistic interpretation of quantum mechanics.\\

\subsection{Approximate solutions of the Schr\"odinger equation}\label{Four_one}

\noindent First, one can argue that the validity of quantum dynamics need not be associated with the exact solvability of the Schr\"odinger equation. Accordingly, the relevance of quantum mechanical description to a particular system cannot be decided based only on the availability (or nonavailability) of the exact solutions to the Schr\"odinger equation of this system.\\

\noindent Yes, it is certainly true that many features of quantum mechanics can be drawn from the approximate solutions to the Schr\"odinger equation (e.g., in the limit when these solutions assume a semiclassical form), but even so the applicability of quantum mechanics \textit{as a matter of principle} should be discussed only using the exact solutions.\\

\noindent This is so because the exact solutions to the Schr\"odinger equation play an extremely important role in quantum physics \cite{Ushveridze}. Apart from being a training ground for elaborating various approximate and qualitative methods and used as zeroth-order approximations for constructing various perturbative schemes, these solutions are the only ones that describe \textit{actual physical reality} (given, of course, that the Schr\"odinger equation describes physical reality). Consequently, only such solutions should be used for testing foundational hypotheses and assumptions of quantum mechanics.\\

\noindent Otherwise, i.e., in case approximate solutions of the Schr\"odinger equation are used for such a purpose, this could lead to a conceptual confusion.\\

\noindent For example, asserting that quantum formalism is applicable even to Hyperion (one of the moons of Saturn) and at the same time considering only approximate solutions to Schr\"odinger’s equation governing the dynamics of this planetary object (e.g., the WKB approximation), one immediately realizes that the approximate quantum description of Hyperion’s rotation will break down very soon: As a result of chaotic evolution, after a short period of time Hyperion has to be (in accordance with the given description) in an extremely non-classical state of rotation, in contrast to what is observed (see details of this description in \cite{Zurek}).\\

\noindent So, to explain away such a discrepancy, the mechanism of decoherence is called as the practically irreversible and practically unavoidable disappearance of certain phase relations from the quantum state of Hyperion by interaction with its environment according to the Schr\"odinger equation. But then again, this mechanism is asserted without computational evidence demonstrating the truth of the proposition that the quantum state of Hyperion \textit{exists in the first place}, i.e., without finding first the set of all exact solutions of the Schr\"odinger equation for Hyperion.\\

\noindent Meanwhile, if one were to accept that there is no algorithm for finding the solution set of the Schr\"odinger equation for an arbitrary system, then one would agree that the proposition of the existence of the quantum state of Hyperion has no proof. Hence, this proposition together with the resulting conclusion of the loss of quantum predictability of Hyperion's movement are neither true nor false, which implies that treating Hyperion quantum mechanically has no meaning.$\,$\footnote{\label{f26}By contrast, for some elementary (or `toy') systems, for which the expressions for quantum states have been demonstrated explicitly, taking into account the environment can provide a reliable explanation for the quantum-to-classical transition (see, for example, paper \cite{Cucchietti} and review \cite{Schlosshauer14}).\vspace{5pt}}\\

\subsection{Super-Turing computability}\label{Four_two}

\noindent Second, one can object that the provability of the statement ${\Pi}_{\mathit{u}}(H_S)$ is controlled by its Turing-computability. For admitting models of computation known as hypercomputers or super-Turing machines, which have capabilities beyond those of a Turing machine -- classical and quantum alike, the decision problem of the Schr\"odinger equation can be made answerable universally, even for systems whose Hilbert spaces are infinite-dimensional.\\

\noindent For example, one of the hypercomputers called an ``infinitely accelerating Turing machine'' can perform infinite amount of elementary operations in a finite amount of time by executing those operations in infinitely short times \cite{Copeland}. In doing so, the accelerating Turing machine can prove the statement ${\Pi}_{\mathit{u}}(H_S)$ for any system $S$ merely by brute force (since in that case a brute force method would halt in a finite time). This means that super-Turing computability might eliminate the Heisenberg's cut by rendering quantum dynamics universal (and the assertion of the existence of the wave function of the entire universe truthful) even from the constructivist point of view.\\

\noindent However, none of the proposed so far models of hypercomputation seems to be physically constructible and reliable.$\,$\footnote{\label{f27}For details see the paper \cite{Davis} that discusses the possibility of computation-like processes which transcend the limits imposed by the Church–Turing hypothesis.\vspace{5pt}} And what is more, any useable physical representation of a Turing-uncomputable function would imply evidence that the universe is not as we imagine it. For example, a physical realization of the infinitely accelerating Turing machine would mean that there exist time intervals shorter than the level of the Planck time widely considered as the scale at which current physical theories fail.\\

\noindent Therefore, even an allowance for the hypothesis of hypercomputation cannot change the fact that in the present state of the knowledge the statement ${\Pi}_{\mathit{u}}(H_S)$ is undecidable for an arbitrary physical system.\\

\subsection{The many-worlds interpretation}\label{Four_three}

\noindent Third, one can argue that the applicableness of quantum mechanics should rely not on the constructibility (computability) of its mathematical entities, but rather on \textit{the elegance of its first principles}. Along these lines, it is unimportant whether or not the quantum states could be actually computed because their existence is stipulated by the basic principles of quantum theory.\\

\noindent For instance, the MWI theory (also referred to as the many-worlds interpretation) asserts the objective reality of the wave function of the universe. Accordingly, in this theory the truthfulness of the existence of the universal wavefunction does not depend on mathematical techniques of solving the Schr\"odinger equation for the universe but simply follows from the postulate. As a consequence, the law of excluded middle is obligatory in the many-worlds interpretation since every statement about the universal wavefunction -- i.e., about the objective reality -- must be either true or false.\\

\noindent An argument like this is logically equivalent to the claim that in physics there are experimentally testable predictions, which are not computable from the data.\\

\noindent Indeed, because in the MWI theory the existence of the universal wave function is accepted without requiring that it be established by a constructive proof (e.g., computational evidence), it indicates that this function is not a necessarily computable one \cite{Vaidman, Arve}. In other words, it is not required that the universal wave function must be a quantity or an expression obtained by actual calculation from the Schr\"odinger equation. On the other hand, the many-worlds interpretation is presented (at least by its proponents) not as a metaphysical or philosophical conjecture but as a physical theory which has the ability (or capacity) for testable predictions. So, as follows, this theory accepts verifiable predictions about the universe that are not computed from the data about the universe, that is, from the Schr\"odinger equation describing all the physical objects in the universe.\\

\noindent The problem here is not that such uncomputable yet verifiable predictions are not possible. The problem is that in the MWI theory is not clear when the quantum states can be actually computed and when they become postulated or, otherwise stated, when the Schr\"odinger equation is a tool for calculation of physical properties and when it and its solutions turn into utter symbols. Under the many-worlds interpretation, the Schr\"odinger equation must hold true all the time everywhere. So, why does the meaning of this equation change?\\

\noindent To be sure, applying the Schr\"odinger equation to the ``smallest possible'' quantum system consisting of a pair of 1-qubit systems -- the microscopic measured system $S$ and the microscopic ``observer'' $M$, one can actually solve the wave equation and prove the existence of the combined observer–object's wavefunction (e.g., by explicitly demonstrating it). Clearly, in such a case the Schr\"odinger equation does not differ much from any other (differential) equations known in physics.\\

\noindent While on the contrary, applying the wave equation to the system comprising the microscopic object $S$ and the really macroscopic ``observer'' $M$ (such as the universe), one must postulate the existence of the combined observer–object's wavefunction instead of proving it. Evidently, in this case, the Schr\"odinger equation is no longer a physical equation but just a symbol representing another symbol -- the wavefunction of the universe entangled with the measured system $S$. So, the question is, when does the transformation of this equation take place?\\

\noindent Even assuming that one is ready to accept that the Schr\"odinger equation changes its meaning during the interaction of $S$ and $M$ (that starts in the microscopic world when the number of degrees of freedom of the observer $M$ involved in the interaction is still small and goes further into the macroscopic world involving more and more degrees of freedom of $M$), hardly such a change can be portrayed as an ``elegance of the first principles''.\\

\subsection{Empiricism of logic}\label{Four_four}

\noindent Finally, one may say that decidability (or computability) as a question of the existence of an effective method for solving a problem is not an attribute or quality of material, physical objects; therefore, it cannot be a rule or principle of a physical theory. Likewise, assumptions of intuitionistic logic as a creation of the mind cannot have any connection with foundations of quantum mechanics.\\

\noindent Actually, this is an open question. Theoretically, the principles of logic might be susceptible to revision on empirical grounds. As it is argued in the celebrated works \cite{Putnam, Dummett}, logic may be empirical, that is, the laws of logic may, or should, be empirically determined.$\,$\footnote{\label{f28}See also review \cite{Bacciagaluppi} on the same topic.\vspace{5pt}}\\

\noindent In accordance with this idea, in the present paper the decision problem of the Schr\"odinger equation is symbolized as a logical statement so that the validity of ``quantum fundamentalism'' (asserting that we are living in a quantum world or, in other words, that everything in the universe – including the universe itself – is ultimately describable in quantum-mechanical terms)$\,$\footnote{\label{f29}See the paper \cite{Zinkernagel15} discussing at length this principle and Bohr's position regarding it.\vspace{5pt}} is provable from this statement. As it is demonstrated, the law of excluded middle would be applicable to the introduced statement if and only if quantum fundamentalism held.\\

\noindent On the other hand, because the decision problem of the Schr\"odinger equation is in general undecidable, such a statement is allowed to be other than true or false, explicitly, it may fail to have truth values at all. This provides a compelling case for abandoning the law of excluded middle together with quantum fundamentalism.\\

\bibliographystyle{References}
\bibliography{References}

\end{document}